\documentclass[12pt]{article}
\usepackage{amsmath,amsfonts,amssymb}
\usepackage[nosort]{cite}
\usepackage{hyperref}
\usepackage{graphicx}
\usepackage{color}
\usepackage{epsfig}
\usepackage{psfrag}	
\usepackage{tikz}
\usepackage{slashed}
\usepackage{ulem}
\usepackage{comment}
\usepackage[all]{xy}

\usepackage{color}

\newcommand \be{\begin{eqnarray}}
\newcommand \ee{\end{eqnarray}}

\makeatletter\@addtoreset{equation}{section}\makeatother

\DeclareMathOperator{\Tr}{Tr}

\let\Re\undefined
\DeclareMathOperator{\Re}{Re}
\let\Im\undefined
\DeclareMathOperator{\Im}{Im}

\DeclareMathOperator{\sign}{sign}

\def\bR {\mathbb{R}}

\newcommand{\bea}{\begin{eqnarray}}
\newcommand{\eea}{\end{eqnarray}}
\newcommand{\beq}{\begin{equation}}
\newcommand{\eeq}{\end{equation}}
\newcommand{\bal}{\begin{equation}\begin{aligned}}
\newcommand{\eal}{\end{aligned} \end{equation}}

\newcommand{\vev}[1]{{\left< {#1} \right>}}

\newcommand{\eqn}[1]{(\ref{#1})}

\newcommand{\address}[1]{\vbox{\center\em#1}}
\renewcommand{\title}[1]{\vbox{\center\huge{#1}}\vspace{5mm}}

\newcommand{\cL}{{\mathcal L}}

\newcommand{\cN}{{\mathcal N}}

\newcommand{\cO}{{\mathcal O}}

\begin{document}
\bibliographystyle{utphys2}

\begin{titlepage}
\rightline{Imperial-TP-EV-2018-01}
\rightline{NORDITA 2018-120}
\begin{center}

\vspace{5mm}

\title{Deformations of the circular Wilson loop \\and spectral (in)dependence}

\vspace{4mm}

\renewcommand{\thefootnote}{$\alph{footnote}$}

Michael Cooke,$^1$\footnote{\href{mailto:cookepm@tcd.ie}
{\tt cookepm@tcd.ie}}
Amit Dekel,$^2$\footnote{\href{mailto:amit.dekel@nordita.org}
{\tt amit.dekel@nordita.org}}
Nadav Drukker,$^1$\footnote{\href{mailto:nadav.drukker@gmail.com}
{\tt nadav.drukker@gmail.com}}\\
Diego Trancanelli,$^3$\footnote{\href{mailto:dtrancan@if.usp.br}
{\tt dtrancan@if.usp.br}}
and
Edoardo Vescovi$^{3,4}$\footnote{\href{mailto: e.vescovi@imperial.ac.uk}
{\tt e.vescovi@imperial.ac.uk}}
\vskip2mm
\address{
$^1$Department of Mathematics, King's College London,
\\
The Strand, WC2R 2LS London, United Kingdom
}
\address{
$^2$
Nordita, KTH Royal Institute of Technology and Stockholm University,\\
Roslagstullsbacken 23, SE-106 91 Stockholm, Sweden
}
\address{
$^3$Institute of Physics, University of S\~ao Paulo,
05314-970 S\~ao Paulo, Brazil
}
\address{
$^4$The Blackett Laboratory, Imperial College, London SW7 2AZ, United Kingdom
}

\renewcommand{\thefootnote}{\arabic{footnote}}
\setcounter{footnote}{0}

\end{center}

\vspace{1mm}
\abstract{
\normalsize{
\noindent
In this paper we study the expectation value of deformations of the circular Wilson loop in  $\cN=4$ 
super Yang-Mills theory. The leading order deformation, known as the Bremsstrahlung function, 
can be obtained exactly from supersymmetric localization,  so our focus is on deformations at higher 
orders. We find  simple expressions for the expectation values for generic deformations at the quartic 
order  at one-loop at weak coupling and at leading order at strong coupling. We also present a  very 
simple  algorithm (not requiring integration) to evaluate the two-loop result.  We find that an exact 
symmetry of the strong coupling  sigma-model, known as the spectral-parameter  independence, is 
an approximate symmetry at weak coupling, modifying the expectation value starting only at the 
sextic order in the deformation. Furthermore, we find very simple patterns for how the spectral 
parameter can  appear in the weak coupling calculation, suggesting all-order structures.
}}
\vfill

\end{titlepage}



\section{Introduction}
\label{sec:intro}

Wilson loop operators are among the most interesting observables that can be studied in gauge theories. 
For ${\cal N}=4$ super Yang-Mills (SYM) theory in four dimensions, these operators are supported along a contour in 
spacetime and may couple to the scalar fields $\Phi^I$ in the gauge multiplet. 
The most symmetric (and supersymmetric) operator is the $1/2$ BPS circle
\beq
W = \frac{1}{N} \Tr {\cal P} \exp \oint d\theta \left(i A_\mu \dot{x}^\mu(\theta) + |\dot x|\Phi^1\right),
\qquad x^\mu(\theta)=(\cos \theta,\sin \theta,0,0)\,.
\label{couplings}
\eeq
The exact vacuum expectation value (VEV) of the $1/2$ BPS circle is given by Laguerre polynomials, 
which reduce to a Bessel function in the large $N$ limit
\cite{erickson,gross,pestun}
\beq
\langle W\rangle 
=\frac{1}{N}L_{N-1}^1(\lambda/4N)e^{\lambda/8N}
= \frac{2}{\sqrt{\lambda}}I_1\big(\sqrt\lambda\big)+\cO(1/N^2)\,.
\label{circleVEV}
\eeq

This paper is concerned with non-supersymmetric deformations of this operator, where the contour 
is still restricted to be inside the same Euclidean plane of the circle and with the same constant 
scalar coupling. The contour is represented by the function (and Fourier series)
\beq
\label{fourier}
X(\theta)=x^1(\theta)+ix^2(\theta)
=e^{i\theta+g(\theta)}\,,\qquad
g(\theta)=\sum_{n=-\infty}^\infty b_ne^{i n\theta}\,.
\eeq
Clearly, for $g(\theta)=0$ we recover the circle, so small deformations can be written as a power series in $b_n$. 
For infinitesimal $b_n$'s, neither self-intersections nor cusps form, and, consequently, no 
divergences appear in the evaluation of the expectation values. We restrict our attention to the planar limit, 
mainly to allow for comparisons with the strong coupling results of \cite{Ishizeki:2011bf}.

The order $\cO(b_n^2)$ has already been studied in the past and is known as the wavy-circle (or wavy-line) 
approximation \cite{Semenoff:2004qr}. It turns out that, at this order, the coupling dependence of the expectation 
value of these Wilson loops is given by the Bremsstrahlung function $B(\lambda)$ introduced in \cite{correa}. 
Evaluating the one-loop order in $\lambda$ (or the classical strong coupling limit, using holography) is 
sufficient to determine this expectation value to all orders in perturbation theory \cite{Fiol:2012sg}
\beq
\label{ep2-intro}
\frac{\langle W\rangle |_{b^2}}{\langle W\rangle |_{b^0}}
=\pi^2B(\lambda)\sum_{n=2}^\infty n(n^2-1)|b_n+\bar b_{-n}|^2,
\qquad
B(\lambda)=\frac{1}{4\pi^2}\frac{\sqrt\lambda I_2(\sqrt\lambda)}{I_1(\sqrt\lambda)}\,,
\eeq
where $\langle W\rangle |_{b^0}$ is the expression in (\ref{circleVEV}). 
This vanishes for $b_{-n}=-\bar b_n$, which corresponds to the case of an imaginary $g(\theta)$ and is just 
a reparametrization of the curve. Assuming $b_{-n}=\bar b_n$ and using (\ref{circleVEV}), this can be written as
\beq
\label{ep2-intro2}
\langle W\rangle |_{b^2}
=2I_2(\sqrt\lambda)\sum_{n=2}^\infty n(n^2-1)|b_n|^2.\,
\eeq
These formulas have been previously written as integrals over $g(\theta)$, and this is their 
Fourier representation. This transformation is done explicitly in the one-loop calculation in Section~\ref{sec:one} below.
Indeed, some of the main calculations that we undertake in this paper are to evaluate one- and two-loop integrals in 
the Fourier basis, which is a natural representation of smooth deformations of the circular Wilson loop. 

Much less is known about the corrections beyond the wavy approximation. Going to 
higher orders in $b_n$, one can see that there is no contribution at cubic order thanks to conformal symmetry. 
The first nontrivial order is therefore quartic in $b_n$, which is the focus of this paper. We find closed-form 
expressions at this deformation order for both one-loop in perturbation theory and leading classical order 
at strong coupling. We also present an efficient algorithm to calculate the exact $\cO(\lambda^2)$ result for 
any loop specified by a finite set of non-zero $b_n$'s.

Note that there are several natural other ways to represent deformations of the circular Wilson loop. One is in terms 
of local ``bumps", which can be reinterpreted as the insertion of local operators into the circle and were the subject of 
recent analyses in \cite{Cooke:2017qgm, Giombi:2017cqn, Kim:2017phs, Kim:2017sju}. Another parametrization 
arises in the evaluation of the expectation value of Wilson loops at strong coupling, via holography. It is 
well-known that this evaluation consists in finding the (regularized) area of a minimal string surface anchored 
to the Wilson loop's contour on the $AdS_5$ boundary. In the case of Wilson loops confined to a plane, 
the dual string world-sheet is contained in an Euclidean $AdS_3$ and there is a standard method, 
reviewed in the next section, to find such surfaces through the Pohlmeyer reduction of the bosonic string 
sigma-model \cite{Pohlmeyer:1975nb,Ishizeki:2011bf}. In this representation, the string 
world-sheet (and hence the boundary contour) is determined by a holomorphic function $f(z)$. For the 
$1/2$ BPS circle, this function vanishes, so one can parametrize deformations of the circle in terms of%
\footnote{We adopt this peculiar mismatch between the subscript of $a$ and the power of $z$, as it 
makes many subsequent formulas significantly simpler.}
\beq
f(z)=\epsilon\sum_{p=2}^\infty a_{p}z^{p-2}\,,
\label{fofz}
\eeq
with small $\epsilon$. The map between the two descriptions (\ref{fourier}) and (\ref{fofz}) was studied in great 
detail in \cite{Dekel:2015bla} and is reviewed below in Section~\ref{sec:contours}. In that paper, the map was 
established between particular examples of deformations of the circle and the function $f(z)$ to high orders 
in the $\epsilon$-expansion. Here we consider the most general deformation, but focus mainly on the order 
$\epsilon^4$. 
As reviewed below, the map between the two descriptions, albeit complicated, is explicit. Since this map is non-linear, 
results at a certain order in the coefficients $b_n$ in \eqn{fourier} end up including higher powers of $\epsilon$.

One interesting property of this last representation of the Wilson loops is that the classical action of the 
string does not depend on the overall phase of $f(z)$, so it is invariant under $f(z)\to e^{i\varphi}f(z)$.%
\footnote{One can choose the phase by taking $f(1)$ to be real or the first nonzero $a_p$ to be real, 
but fixing such a choice is not really necessary.}
Indeed 
$\varphi$ is the \emph{spectral parameter} in the integrability description of the string sigma-model and parametrizes 
the ``master" symmetry used to construct an infinite number of non-local Yangian charges in 
\cite{Klose:2016uur, Klose:2016qfv}.

One of the main motivations for this work is to address a puzzle raised in \cite{Dekel:2015bla}.%
\footnote{The spectral parameter was called $\lambda=e^{i\varphi}$ in \cite{Dekel:2015bla} and 
the corresponding transformations were referred to as $\lambda$-deformations. We do not use 
this terminology here to avoid potential confusion with the 't Hooft coupling.} 
The shape of the string world-sheet and of the Wilson loop contour depends on $\varphi$, and indeed 
the expectation value of the Wilson loop may depend on $\varphi$ beyond the classical string limit, 
and in particular at weak coupling in the gauge theory. In the examples studied in \cite{Dekel:2015bla}, 
there was indeed such dependence on $\varphi$ at one-loop order in $\lambda$. This dependence, 
however, did not appear at orders $\epsilon^4$ and $\epsilon^6$ in the deformation around the 
circle. This dependence appeared only at higher orders in $\epsilon$ (either $\epsilon^8$ or 
$\epsilon^{16}$ depending on the specific curve), showing that spectral transformations are 
an unexpected approximate symmetry at weak coupling. The main goal of this paper is to make this 
previous analysis more systematic and extend the weak-coupling computation to two-loops. 
We indeed confirm that there is no $\varphi$-dependence at order $\epsilon^4$ in the one-loop 
expectation value of deformed circles. The same conclusion holds at two-loops for any $f(z)$ polynomials 
of degree equal or less than 10 and we conjecture that it is true for arbitrarily high degree. 
By performing a more comprehensive survey, we find that generically there is 
$\varphi$-dependence at order $\epsilon^6$ in the one-loop expectation value. The absence of 
this dependence in \cite{Dekel:2015bla} is particular to the examples studied there.

In proving all this, we set up an algorithm computing the Wilson loop's expectation value on a deformed circle 
at one- and two-loops in $\lambda$ and up to order $\epsilon^4$. The output is a polynomial of Fourier coefficients 
$b_{n}$ in \eqref{fourier} or of the Taylor coefficients $a_p$ in \eqref{fofz}. The one-loop expression is 
in closed form and can be generalized to higher orders in $\epsilon$. The two-loop results are expressed as 
recurrence relations. The calculation does not involve numerical approximations and it is more efficient 
than evaluating multiple integrals \cite{seminara} when the loop is known in Fourier series representation 
around the circle \eqref{fourier} or in terms of $f(z)$.

The paper is organized as follows. In Section~\ref{sec:contours} we summarize the approach of 
\cite{Ishizeki:2011bf} for constructing minimal surfaces in Euclidean $AdS_3$ and their boundary 
curves from $f(z)$ and focus in Section~\ref{sec:perturbation} on 
small deformations of the circular curve. Section~\ref{sec:VEV} contains the main result of this paper: 
The one-loop and two-loop expectation values at order $\epsilon^2$ and $\epsilon^4$ and 
the spectral independence to order $\epsilon^4$. In Section~\ref{sec:ep6} we discuss the 
spectral dependence at order $\epsilon^6$ and beyond. We comment on interesting patterns in the 
expectation value at classical order at strong coupling in Section~\ref{sec:strong}. We present some 
conclusions and discussions in Section~\ref{sec:discussion}. 
Details on the perturbative algorithm are relegated to an appendix.


\section{From world-sheet to Wilson loop}
\label{sec:contours}

In this section we recount the construction of Wilson loops from the Pohlmeyer description of Euclidean $AdS_3$ 
\cite{Kruczenski:2014bla, Ishizeki:2011bf}, discuss the spectral parameter deformation, 
and focus on the case of nearly circular Wilson loops.

For the string world-sheet we take the unit disk with complex coordinate $z=r e^{i\theta}$. The three-dimensional 
target space can be parametrized by a complex $X$ and a real $Z$, such that the boundary of $AdS$ is at 
$Z=0$, which is also the boundary of the world-sheet, $r=1$. So, at $r=1$ we have a curve $X(\theta)$, 
which is the boundary curve of the string, \emph{i.e.}, the Wilson loop contour.

Using the Pohlmeyer reduction of the sigma-model, a solution of the string equations of motion is 
characterized by a holomorphic function $f(z)$. From it, one constructs a real function $\alpha(z,\bar z)$ 
solving the generalized cosh-Gordon equation
\beq
\label{GcG}
\partial \bar \partial \alpha(z,\bar z) = e^{2\alpha(z,\bar z)} + |f(z)|^2 e^{-2\alpha(z,\bar z)}\,.
\eeq
The regularized area of the world-sheet, which equals the expectation value of the Wilson loop at 
strong coupling, then evaluates to
\beq\label{area}
A_{\text{reg}} = -2\pi\sqrt\lambda - 2\sqrt\lambda \int_{|z|\leq1} d z d\bar z \, |f(z)|^2 e^{-2\alpha(z,\bar z)}\,.
\eeq
Once $f$ and $\alpha$ are given, the world-sheet embedding in target space follows from the solution of 
an auxiliary linear differential equation. Note that both the equation for $\alpha$ and the action depend only 
on the modulus of $f(z)$, so are invariant under $f(z)\to e^{i \varphi} f(z)$. The full string solution, and 
the boundary curve $X(\theta)$, may depend on $\varphi$.%
\footnote{One can view $\varphi$ also as an integration constant arising in the solution of the auxiliary problem.}
This construction therefore leads to a one-parameter family of curves with the same strong coupling 
expectation value. We refer to $\varphi$ as the spectral parameter, and changing the curve induced by it 
as the spectral parameter deformation.

Let us provide further detail on how to extract the contour, given $f(z)$. 
As  shown in \cite{Kruczenski:2014bla}, the behaviour of $\alpha(z,\bar{z})$ close to the boundary 
takes the form
\begin{align}\label{beta2}
\alpha(z,\bar{z}) = -\ln\xi + \xi^2(1+\xi)\beta_2(\theta) + \cO(\xi^4)\,, \qquad \xi = 1 - r^2\,,
\end{align}
with some real function $\beta_2(\theta)$. One can then use \eqref{GcG} to fix the higher order terms 
in the expansion in terms of $\beta_2(\theta)$ and $f(e^{i \theta})$. Examining the full string world-sheet 
near the boundary one can find an equation relating the Schwarzian derivative of the boundary contour 
$X(\theta)$ to the boundary values $\beta_2(\theta)$, $f(e^{i\theta})$ as well as $e^{i\varphi}$:
\begin{equation}
\label{eq:SchwarzDerEq}
\begin{split}
\{X(\theta),\theta\}
\equiv\frac{X'''(\theta)}{X'(\theta) }
-\frac{3}{2}\left(\frac{X''(\theta)}{X'(\theta)}\right)^2
=
-2V(\theta)\,,
\\
V(\theta)=-\frac{1}{4}+6\beta_2(\theta) +2 i \Im(e^{2 i\theta} e^{i \varphi} f(e^{i\theta}))\,.
\end{split}
\end{equation}
Note that \eqref{eq:SchwarzDerEq} depends on the parametrization of $X(\theta)$ and it holds only for the 
parametrization arising in this construction, where $\theta$ is the polar angle in the world-sheet disk. If 
one starts instead with an arbitrarily parametrized curve $X(s(\theta))$, one would need to follow the 
procedure described in \cite{Kruczenski:2014bla,He:2017zsk} to find the ``correct'' parametrization related 
to the world-sheet description.

An alternative formulation of the problem is in terms of a Schr{\"o}dinger problem. The requirement that the 
surface closes up smoothly inside the disk yields a consistency condition that fixes $\beta_2(\theta)$. 
One needs to tune $\beta_2(\theta)$ so that the solution of the Schr\"odinger equation
\beq
-\chi''(\theta) + V(\theta) \chi(\theta) = 0
\eeq
is anti-periodic in $\theta$ with period $2\pi$ for any $\varphi$. We can also define the periodic function 
$\kappa(\theta)\equiv e^{-i\theta/2}\chi(\theta)$, satisfying
\beq
\label{sch2}
-\kappa''(\theta)-i\kappa'(\theta)+\left(V(\theta)+\frac{1}{4}\right)\kappa(\theta)=0\,.
\eeq
This alternative is purely algebraic and one has not to solve the cosh-Gordon equation. It does introduce a 
complication that to fully determine the solution at a fixed order in $\epsilon$ one has to include terms 
of higher orders, so it is not a very efficient algorithm.

Now we see why it is advantageous in this formalism to define the Wilson loop in terms 
of $f(z)$ rather than by the boundary contour $X(\theta)$ (assuming of course that we 
are interested in general contours, rather than a specific shape). 
First, a boundary contour $X(\theta)$ is not invariant under conformal 
transformations, which complicates a general classification of curves, whereas $f(z)$ 
is invariant. Second, the spectral parameter deformation is natural when starting with $f(z)$, and 
then solving for the boundary curve $X(\theta)$. The inverse procedure, going from $X(\theta)$ to 
$f(z)$ and $\varphi$, requires also to reparametrize the curve in terms of the correct angle in the 
unit disc \cite{Kruczenski:2014bla}.%
\footnote{An analytic solution for wavy circles was developed in \cite{Dekel:2015bla} and a numerical 
approach for generic curves in \cite{He:2017zsk}, but both are complicated.}
Lastly, turning on $\varphi$ by a spectral deformation, which modifies $f(z)$ by a constant phase, 
influences one part of the the calculation outlined above.


\subsection{Perturbations around the circle}
\label{sec:perturbation}

Our starting point is the circle with $f(z)=0$. We label small deformations by $f(z)$ proportional to a 
small parameter $\epsilon$ and by their Taylor expansion around the origin as in \eqn{fofz}. Clearly 
the circle is invariant under the spectral deformation 
$f(z)\to e^{i\varphi}f(z)$. The next simplest case is when $f(z)$ is a monomial, namely 
$f(z) = \epsilon \, a_p z^{p-2}$. The spectral deformation shifts the phase of $a_p$. The resulting generalized 
cosh-Gordon equation \eqref{GcG} is independent of $\theta$, which means $\alpha(r,\theta) = \alpha(r)$, 
and equation \eqref{beta2} implies that $\beta_2$ is a constant. Indeed we can absorb $\varphi$ in $z$, 
which amounts to a shift of $\theta$. Therefore, the spectral deformation is equivalent to rotations of the 
curve $X(\theta)$ and the VEV of the Wilson loop does not change. These choices of $f(z)$ were studied 
in \cite{Huang:2016atz} and the exact regularized area was found. The one-loop and two-loop expressions for 
such loops can be extracted from the expressions in subsequent sections.

We now turn to study deformations with general $f(z)$. 
The $\epsilon$-expansion of the Wilson loop's expectation value involves only 
even powers of $\epsilon$. Indeed flipping $f(z) \to -f(z)$ induces $X\to X^*$ in 
\eqref{eq:SchwarzDerEq}, which is a reflection in the $(x^1,x^2)$ plane. 
Note that the VEV of the Wilson loop at order $\epsilon^{2l}$ is a degree $2l$ 
polynomial in the Taylor coefficients $a_p$ and $\bar a_p$. So to this order it suffices 
to consider $f(z)$ which is a general polynomial with up to $2l$ nonzero $a_p$. 

The fact that there are no terms at order $\epsilon$ implies that expanding $X(\theta)$ to order 
$\epsilon$ contributes to the final answer only at order $\epsilon^2$. More generally, any 
perturbative calculation (like \eqn{1loop} and \eqn{gen_2} below) vanishes when one $X(\theta_i)$ is 
a deformation of the circle at order $\epsilon^{2l}$ and all other $X(\theta_j)$ are undeformed circles. 
We therefore require the expression for the curve $X(\theta_i)$ at order $\epsilon^{2l-1}$.

We tackle the problem by expanding $X(\theta)$, $\alpha(z,\bar{z})$ and $\beta_2(\theta)$ in power 
series in $\epsilon$, which allows to solve the nonlinear equation \eqref{GcG} and invert the 
Schwarzian derivative equation \eqref{eq:SchwarzDerEq} perturbatively. The starting point is
\begin{align}
\label{alpha}
\alpha(z,\bar{z})&=\sum_{l=0}^\infty \epsilon^{2l}\alpha^{(2l)}(z,\bar{z})\,,
\hskip-2cm&
\alpha^{(0)}(z,\bar{z})&=-\log(1-r^2)\,,\\
\beta_2(\theta)&=\sum_{l=0}^\infty \epsilon^{2l}\beta^{(2l)}_2(\theta)\,,
\hskip-2cm&
\beta^{(0)}_2(\theta)&=0\,,
\end{align}
where the lowest order terms match those for the circle. 
Plugging \eqref{alpha} and \eqref{fofz} into \eqn{GcG} gives differential recursion relations for $\alpha^{(2l)}$. 
The first three are
\bal\label{eq_a2}
\cL \alpha^{(2)} 
&= 4(1-r^2)^2|f/\epsilon|^2\,,
\qquad\qquad\qquad
\cL = \partial_r^2 + \frac{\partial_r}{r} + \frac{\partial^2_\theta}{r^2} - \frac{8}{(1-r^2)^2}\,,
\\
\cL \alpha^{(4)}
&= \frac{8(\alpha^{(2)})^2}{(1-r^2)^2} -8(1-r^2)^2|f/\epsilon|^2 \alpha^{(2)}\,,
\\
\cL \alpha^{(6)} 
&=\frac{16\alpha^{(2)}\big((\alpha^{(2)})^2+3\alpha^{(4)}\big)}{3(1-r^2)^2} 
-8(1-r^2)^2|f/\epsilon|^2\big(\alpha^{(4)} - (\alpha^{(2)})^2\big)\,.
\eal

To determine the VEV at order $\epsilon^4$, we only need to find $\alpha^{(2)}(z,\bar{z})$. 
It is useful to represent the holomorphic function as \cite{Dekel:2015bla}
\begin{gather}\label{f(G)}
f(z) =-\frac{1}{2}\left(z G'''(z) +3 G''(z) \right)\,,
\end{gather}
where clearly
\begin{flalign}\label{G}
G(z) &=-2 \, \epsilon \sum_{p=2}^\infty \frac{a_p z^{p}}{p(p^2-1)}\,.
\end{flalign}
The solution of the linear differential equation for $\alpha^{(2)}$ is then
\bal\label{alpha_hom_inhom}
\alpha^{(2)}(z,\bar{z})&=|W(z,\bar{z})|^2-\frac{z \bar{G}(\bar{z})W(z,\bar{z})+\bar{z} G(z)\bar{W}(z,\bar{z})}{1-|z|^2}\cr
&\quad{}
+\frac{1+|z|^2}{1-|z|^2}(T(z)+\bar{T}(\bar{z}))+z T'(z)+\bar{z} \bar{T}'(\bar{z})\,,
\eal
where the first line is a particular solution to the inhomogeneous equation with
\begin{flalign}
\label{W}
W(z,\bar{z}) &\equiv \frac{1}{2}z(|z|^2-1)G''(z)-G'(z)\,.
\end{flalign}
The second line is a general solution to the homogeneous equation, which is fixed by imposing 
the boundary condition that there are no poles for $\xi\to0$ in \eqref{beta2}. The result is
\begin{flalign}\label{T}
T(z) &= -2 \epsilon^2 \sum_{p=2}^\infty\frac{|a_p|^2}{p^2(p^2-1)^2}
-2 \epsilon^2 \sum_{p=2}^\infty \sum_{q=p+1}^\infty 
\frac{(p+q) z^{q-p}a_q \bar{a}_p}{p(p^2-1)q(q^2-1)}\,.
\end{flalign}

Equipped with $\alpha^{(2)}(z,\bar{z})$, we can read off $\beta_2(\theta)$ in two equivalent ways. 
Using \eqref{beta2} and expanding $\alpha^{(2)}$ to order $\xi^2$ gives 
\begin{flalign}\label{b22}
\beta^{(2)}_2(\theta) &=-\frac{1}{3}\sum_{p=2}^\infty \frac{|a_p|^2}{p^2-1}\,.
\end{flalign}

In the alternative approach, based on the periodicity of $\kappa$, 
one can expand $\kappa(\theta) = \sum_{l=0}^\infty \epsilon^l \kappa^{(l)}(\theta)$ 
with $\kappa^{(0)}(\theta) = 1$ and the Schr\"odinger equation \eqn{sch2} becomes
\bal
L_1 \kappa^{(1)} = & 2 i \Im(e^{2 i\theta} e^{i \varphi} f(e^{i\theta}))\,,
\qquad\qquad\qquad\qquad
L_1\kappa\equiv \kappa'' + i \kappa'\,,\\
L_1 \kappa^{(2)} = & 2 i \Im(e^{2 i\theta} e^{i \varphi} f(e^{i\theta})) \kappa^{(1)} + 6\beta_2^{(2)}\,,\\
L_1 \kappa^{(3)} = & 2 i \Im(e^{2 i\theta} e^{i \varphi} f(e^{i\theta})) \kappa^{(2)} + 6\beta_2^{(2)}\kappa^{(1)}\,,\\
L_1 \kappa^{(4)} = & 2 i \Im(e^{2 i\theta} e^{i \varphi} f(e^{i\theta})) \kappa^{(3)} + 6\beta_2^{(2)}\kappa^{(2)} + 6\beta_2^{(4)}\,,
\eal
and so on. The inverse of $L_1$ on a basis of exponentials is given by
\begin{align}
L_1^{-1} e^{i k \theta} =
\begin{cases}
\displaystyle-\frac{e^{i k \theta}}{k(k+1)}\,,\quad & k\neq 0,-1\,,
\\[6pt]\displaystyle
-i \theta\,, & k=0\,, \\
\displaystyle
e^{-i \theta } \left(1+i\theta\right)\,, & k=-1\,,
\end{cases}
\end{align}
and that allows to solve for $\kappa^{(l)}$. The kernel is $\text{ker}(L_1) = c_1 + c_2 e^{-i \theta}$, 
so the modes $0$ and $-1$ must not appear on the right-hand sides for the solution to be periodic. 
Indeed there are no such terms in $V$, so we can solve the first line. 
At higher orders, the requirement of periodicity of $\kappa^{(l)}$ imposes conditions on $\beta_2^{(l)}$, 
such that there are no $k=0,-1$ modes on the right-hand side. Unfortunately, this does not 
completely determine $\beta_2^{(l)}$, order by order, and constraints from the higher order terms 
are needed. Still, one can rederive \eqref{b22} above from these requirements.

We can now solve perturbatively for the shape of the curve. We take 
\begin{gather}\label{path_series}
X(\theta) = \exp\left(i\theta+\sum_{l=1}^\infty \epsilon^l g^{(l)}(\theta)\right)
\end{gather}
and plug it into the Schwarzian derivative equation \eqn{eq:SchwarzDerEq}. 
The first few equations are
\bal
\label{L2}
L_2 g^{(1)}&=4\Im\big(e^{2 i \theta} e^{i \varphi}f(e^{i \theta})/\epsilon\big)\,,
\qquad\qquad\qquad
L_2 g \equiv g''' + g'\,,
\\
L_2 g^{(2)}&=\frac{i}{2}(g^{(1)\prime})^2
-\frac{3i}{2}(g^{(1)\prime\prime})^2-ig^{(1)\prime}g^{(1)\prime\prime\prime}-12i\beta_2^{(2)}\,,
\\
L_2 g^{(3)}&=g^{(1)\prime\prime\prime}\left((g^{(1)\prime})^2-ig^{(2)\prime}\right)-3 i g^{(1)\prime\prime} g^{(2)\prime\prime}
+g^{(1)\prime}\left(3 (g^{(1)\prime\prime})^2 +i g^{(2)\prime}-i g^{(2)\prime\prime\prime}\right)\,.
\eal
Now we can use the inverse of $L_2$, which is
\beq\label{L2inv}
L_2^{-1} e^{i k \theta} =\frac{i e^{i k \theta}}{k(k^2 - 1)}\,,\qquad k\neq0,\pm1\,.
\eeq
$L_{2}^{-1}$ acting on $e^{ik\theta}$ with $k=0,\pm1$ would give non-periodic functions, but these powers never 
occur on the right-hand side.

Let us apply the above procedure. Plugging \eqref{fofz} into the first line of \eqn{L2} gives
\beq
L_2 g^{(1)}
=4 \Im\left(e^{2 i \theta}e^{i\varphi}f(e^{i \theta})\right) 
= -2 i\epsilon\sum_{p=2}^\infty\left(a_p e^{ip\theta+i\varphi} 
- \bar a_p e^{-ip\theta-i\varphi}\right).
\eeq
Inverting $L_2$ \eqref{L2inv} gives $X(\theta)$ to linear order in $\epsilon$
\beq
\label{g1}
g^{(1)}(\theta)
=
2\sum_{p=2}^\infty\frac{a_p e^{i(p\theta+\varphi)}+\bar a_p e^{-i(p\theta+ \varphi)}}{p(p^2-1)}\,.
\eeq
The Fourier modes in \eqn{fourier} at this order follow from the comparison with \eqref{path_series}:
\beq
b_n=\begin{cases}
0\,,&n=-1,0,1,
\\[6pt]\displaystyle
\frac{2\epsilon a_{n}e^{i\varphi}}{n(n^2-1)}\,,&n\geq2\,,
\\[12pt]\displaystyle
-\frac{2\epsilon \bar a_{-n}e^{-i\varphi}}{n(n^2-1)}\,,&n\leq-2\,.
\end{cases}
\eeq

Let us record also the next term
\bal
\label{explicit-g}
g^{(2)}(\theta)&=
\sum_{p=2}^\infty
\frac{ p(5 p^2+1) a_{p}^2 e^{2 i (p\theta+\varphi )}
-4(4 p^2-1) |a_{p}|^2
-p(5 p^2+1) \bar{a}_{p}^2 e^{-2 i (p\theta+\varphi )}}{p^2(p^2-1)^2 (4 p^2-1)}
\\&\quad
+
\sum_{p>q\geq2}\Bigg[
\frac{4(p^2+3 p q+q^2+1)\left(a_{p}a_{q} e^{i ((p+q)\theta+2 \varphi )}
-\bar{a}_{p}\bar{a}_{q}e^{-i ((p+q)\theta+2 \varphi)}\right)}
{(p^2-1)(q^2-1)(p+q)((p+q)^2-1)}
\\&\hskip2.4cm
-\frac{4a_{p}\bar{a}_{q} e^{i(p-q)\theta}-4 \bar{a}_{p}a_{q} e^{-i(p-q)\theta}}{p (p^2-1) (q^2-1)}
\Bigg]\,,
\eal
from which we can similarly extract the $\epsilon^2$-correction to $b_{n}$. 
We have calculated also the 
general expression for $g^{(3)}(\theta)$, but have chosen not to write it, as it is 
very cumbersome and we explained in full detail the algorithm to evaluate it.


\section{Expectation value of perturbed circle}
\label{sec:VEV}

We have two representations of the Wilson loop, in terms of the path as in \eqn{fourier} 
and in terms of the holomorphic function $f(z)$ on the string world-sheet. We now 
proceed to evaluate the expectation value of the Wilson loop at one and two loops at 
weak coupling in terms of the Fourier coefficients $b_{n}$ and, alternatively, in terms 
of the Taylor coefficients $a_p$. The map between these coefficients has been derived in the last section.

\subsection{One-loop order}
\label{sec:one}
The one-loop expression for the Wilson loop is given by the double integral
\beq\label{1loop}
\vev{W}_{\lambda}
=
\frac{\lambda}{16\pi^2}\oint d\theta_1\, d\theta_2\, I(\theta_1,\theta_2)\,,
\qquad
I(\theta_1,\theta_2)
=-\frac{\Re(\dot X(\theta_1)\dot{\bar{X}}(\theta_2))
- |\dot X(\theta_1)||\dot X(\theta_2)|}{|X(\theta_1)-X(\theta_2)|^2}.
\eeq
To quadratic order in $g(\theta)$, this is
\begin{align}
I(\theta_1,\theta_2)&=\frac{1}{2}
-\frac{i (e^{i \theta_1}+e^{i \theta_2}) (\dot g(\theta_1)-\dot g(\theta_2))}
{2 (e^{i\theta_1}-e^{i \theta_2})}
\\&\quad
+\frac{e^{i(\theta_1+\theta_2)}}{2 (e^{i \theta_1}-e^{i \theta_2})^2}
\Big[(g(\theta_1)-g(\theta_2))^2
-\dot g(\theta_1)^2-\dot g(\theta_2)^2
+2\cos(\theta_1-\theta_2)\dot g(\theta_1) \dot g(\theta_2)
\Big]\,.\nonumber
\end{align}

For the perturbed circle we plug in the Fourier representation of $g(\theta)$ \eqn{fourier} and expand to 
fixed order in powers of $b_n$, or alternatively $\epsilon$. We end up with a sum of many terms of the form 
\beq
\frac{e^{in_{1}\theta_{1}+in_{2}\theta_{2}}}
{(e^{i\theta_{1}}-e^{i\theta_{2}})^p}\,,
\eeq
with varying $n_1$, $n_2$ and $p$. If we specify a finite set of Fourier modes and combine all the 
terms, the numerator factorizes and cancels the denominator. It is then trivial to do the integral, by picking 
out the zero-mode.

Unfortunately this factorization is not evident when the Fourier numbers are left as arbitrary 
variables $n_i$. There are two approaches to that case, first one can evaluate the result 
for many combinations of $n_i$ and try to fit it to a function of these integers. We chose instead 
a recursive algorithm to calculate the integrals
\beq
\label{def_A}
A_{n_{1},n_{2}}^{p}
=\frac{1}{4\pi^2}\oint d\theta_{1} d\theta_{2} \, 
\frac{e^{in_{1}\theta_{1}+in_{2}\theta_{2}}}
{(e^{i\theta_{1}}-e^{i\theta_{2}})^p}\,.
\eeq
For $p=0$ this is simply $A_{n_{1},n_{2}}^{0}=\delta_{0,n_{1}}\delta_{0,n_{2}}$. 
For $p>0$, instead of performing the integrals, we view them as formal objects 
satisfying the parity condition and recurrence relations, arising from combinations of integrands 
with factorizable numerators
\beq\label{eqs_for_A}
A_{n_{1},n_{2}}^{p}=(-1)^{p}A_{n_{2},n_{1}}^{p}\,,
\qquad
\sum_{k=0}^{p}\binom{p}{k}(-1)^{k}A_{n_{1}+p-k,n_{2}+k}^{p}=A_{n_{1},n_{2}}^{0}\,.
\eeq
For $p=2$ this is solved by
\beq
\label{A2}
A_{n_{1},n_{2}}^{2} 
=\frac{1}{4}|n_{1}-n_{2}|\delta_{2,n_{1}+n_{2}}+C^{(1)}_{n_{1}+n_{2}}\,,
\eeq
with arbitrary $C^{(1)}_n$, which will cancel when we take the sum of integrals evaluating the 
Wilson loop. Note that each of the integrals (for $p>0$) is divergent, but the final expression for the 
Wilson loop at a given loop order is finite, so the arbitrary $C^{(1)}_n$ does in fact contain these divergences.

One can proceed to higher orders in the $g(\theta)$ expansion. At quartic order 
we find
\beq
\label{A4}
A_{n_{1},n_{2}}^{4} 
=\frac{1}{96}|n_{1}-n_{2}|\left((n_{1}-n_{2})^{2}-4\right)\delta_{4,n_{1}+n_{2}}
+C^{(2)}_{n_{1}+n_{2}}(n_{1}-n_{2})^{2}+C^{(3)}_{n_{1}+n_{2}}\,,
\eeq
where again $C^{(2)}$ and $C^{(3)}$ cancel out in the end result. In terms of the Fourier 
coefficients $b_n$, one finds to quartic order
\begin{align}
\vev{W}_{\lambda}
&=\frac{\lambda}{8}
+\frac{\lambda}{4}\sum_{n=2}^\infty n(n^2-1)|b_n|^2
+\frac{\lambda}{6}\sum_{n\geq m\geq l\geq1}\Bigg[
3S^{(1)}_{n,m,l}\times
\\&\quad
\times
l m n(2 l^2+2 m^2+2 n^2+3 l m+3 l n+3 m n-3)
(b_nb_mb_l\bar b_{n+m+l}+\bar b_n\bar b_m\bar b_l b_{n+m+l})
\nonumber\\&\quad
+S^{(2)}_{n,m,l} \,
l(l^4-l^3 m-l^3 n+l^2 m^2-5 l^2 m n+l^2 n^2-6 m^3 n-9 m^2 n^2-6 m n^3
+9 l m^2 n
\nonumber\\&\quad
+9 l m n^2
-4 l^2+l m+l n-m^2+8 m n-n^2+3)
(b_nb_m\bar b_l\bar b_{n+m-l}+\bar b_n\bar b_m b_l b_{n+m-l})
\Bigg],
\nonumber
\end{align}
with the symmetry factors
\beq
\label{Snml}
S^{(1)}_{n,m,l}=\begin{cases}
\frac{1}{6}\,,&n=m=l\,,\\
\frac{1}{2}\,,& n=m\quad\text{or}\quad m=l\,,\\
1\,,&\text{otherwise,}
\end{cases}
\qquad
S^{(2)}_{n,m,l}=\begin{cases}
\frac{1}{8}\,,&n=m=l\,,\\
\frac{1}{2}\,,& n=m\quad\text{or}\quad m=l\,,\\
1\,,&\text{otherwise.}
\end{cases}
\eeq

We can also express this in terms of the coefficients $a_p$ of the function 
$f(z)$. Plugging \eqn{g1}, \eqn{explicit-g} and the order $\epsilon^3$ term
into the last equation, we find
\bal
\label{ep4asnew}
\vev{W}_{\lambda}&=
\frac{\lambda}{8}
+\lambda\epsilon^2
\sum _{p=2}^{\infty }\frac{|a_{p}|^2}{p (p^2-1)}
\\&\quad
+\frac{8}{3}\lambda\epsilon^4\sum _{p\geq q\geq r\geq2}
S^{(2)}_{p,q,r}
\frac{a_{p} \bar a_{q}\bar{a}_{r} a_{q+r-p}
+ \bar{a}_{p} a_{q} a_{r}\bar a_{q+r-p}}
{p (p^2-1) q (q^2-1)r(r^2-1) (q+r)((q+r)^2-1)}
\\&\quad
\times \big[4 p^2(q+r)((q+r)^2-1)+p(-q^4+2 q^3 r+12 q^2 r^2+2 q r^3-r^4+q^2+8 q r+r^2)
\\&\qquad
-(q+r)(2 q^4+3 q^3 r-4 q^2 r^2+3 q r^3+2 r^4+q^2-5 qr+r^2-3)\big]+\cO(\epsilon^6)\,,
\eal
with the same symmetry factor as above.
It is easy to see that the order $\epsilon^2$ term is an alternative representation of 
the coefficient of the Bremsstrahlung function. At order $\epsilon^4$ one could have 
had in principle any polynomial of degree 4 in $a_p$ and $\bar a_p$ with up to four 
different $p$'s, but most of those vanish. Instead, we find that they are all of the 
form $a_{p_1}a_{p_2}\bar a_{p_3}\bar a_{p_4}$, with $p_1+p_2-p_3-p_4=0$. Note 
that all the $\varphi$ dependence in \eqn{explicit-g} completely cancels in this 
expression and it is $\varphi$ independent, confirming the observation in 
\cite{Dekel:2015bla}! As we discuss in Section~\ref{sec:ep6}, 
this independence is generally violated at order $\epsilon^6$ and above.

\subsection{Two-loop order}
\label{sec:two}

A similar analysis can also be done at two-loops order by virtue of a compact 
formula \cite{seminara} which incorporates all two-loop Feynman 
diagrams for a curve in $\bR^2$
\begin{eqnarray}
\label{gen_2}
\vev{W}_{\lambda^2}
&&=
\frac{\lambda^2}{128\pi^4}\oint d\theta_1\,d\theta_2\,d\theta_3\,
\bigg[\varepsilon(\theta_1,\theta_2,\theta_3) I(\theta_1,\theta_3) 
\frac{\Re((X(\theta_3)-X(\theta_2))\dot{\bar{X}}(\theta_2))}{|X(\theta_3)-X(\theta_2)|^2}
\cr
&&\hskip4.5cm 
\times
\log\frac{|X(\theta_1)-X(\theta_2)|^2}{|X(\theta_3)-X(\theta_1)|^2}\bigg]
+\frac{\lambda^2}{2}\left(\frac{1}{16\pi^2}\oint d\theta_1\,d\theta_2 \, I(\theta_1,\theta_2)\right)^2
\cr
&&\quad
-\frac{\lambda^2}{64\pi^4}\int_{\theta_1>\theta_2>\theta_3>\theta_4} 
d\theta_1\,d\theta_2\,d\theta_3\,d\theta_4\,
I(\theta_1,\theta_3)I(\theta_2,\theta_4)\,,
\end{eqnarray}
with $I(\theta_1,\theta_2)$ from \eqn{1loop}. 
The first integral is due to cubic interactions and $\varepsilon$ 
is completely antisymmetric in permutations of its arguments and takes the 
value $1$ for $\theta_1>\theta_2>\theta_3$.

The procedure we employ to evaluate these integrals is a generalization of the one-loop case, 
but it is significantly more complicated and is explained in Appendix~\ref{app:system}. 
Expanding \eqn{gen_2} in powers of the Fourier coefficients $b_n$ gives integrals with all 
the $X(\theta)$ factors in \eqn{gen_2} replaced by exponentials (and arbitrary powers of 
differences of exponentials in the denominator), as in \eqn{def_B} and \eqn{def_C}. 
If we consider the cases of a trivial denominator, the integrals are convergent, 
and these serve again as the boundary conditions for recursion relations for the integrals with 
arbitrary denominator, \eqn{eq_for_B} and \eqn{eq_for_C}. As in the 
one-loop integrals, there are ambiguities in solving the recursion relations which 
are solutions to homogeneous equations, but they drop out from the final expressions.
The complexity of the recursion relations prevented us from finding closed-form solutions like
\eqn{A2} and \eqn{A4}, but for any finite set of Fourier modes with mode 
numbers $n_i$, we can solve the recursion with the aid of a computer.

At order $\epsilon^2$ we know that the result should be determined by the Bremsstrahlung 
function, and in particular should be proportional to the order $\epsilon^2$ term in \eqn{ep4asnew}. 
To verify this, we considered 
\beq
\label{fbinomial}
f(z) = \epsilon\, (a_{p} z^{p-2}+ a_{q} z^{q-2})\,, \quad\text{with}\quad
2 \leq q < p \leq 12\,.
\eeq
We used the algorithm in Appendix~\ref{app:system} to find the result
\beq
\label{2brem}
\vev{W}_{\lambda^2}=
\frac{\lambda^2}{192}+\frac{\lambda^2\epsilon^2}{12}
\sum _{p=2}^{\infty }\frac{|a_{p}|^2}{p (p^2-1)}+\cO(\epsilon^4)\,.
\eeq
The functional form and numerical prefactor indeed agree with the two-loop order of \eqn{ep2-intro2}. 

At order $\epsilon^4$ the result should be quartic in $a_{p_i}$ and $\bar a_{p_i}$, so we need consider $f(z)$ with 
at most four terms 
\beq
f(z) = \epsilon(a_{p_1} z^{p_1-2}+a_{p_2} z^{p_2-2}+a_{p_3} z^{p_3-2}+a_{p_4} z^{p_4-2}). 
\eeq
We computed the result for all such $f(z)$ of degree no larger than 10
\begin{gather}
2 \leq p_4 < p_3 < p_2 < p_1 \leq 12\,.
\end{gather}
We find that of all possible degree-4 monomials in $a_{p_i}$ and $\bar a_{p_i}$, 
the only non-zero contributions come from terms of the form $a_pa_q\bar a_r\bar a_{p+q-r}$, 
with arbitrary $p,q,r$. This fits exactly the pattern at one-loop order \eqref{ep4asnew}.

The fact that the allowed monomials contain an equal number of $a$ and $\bar a$ immediately implies 
that there is no spectral parameter dependence in the final expression. We conjecture that these properties 
extend beyond this range of $p_i$.

Unlike \eqref{ep4asnew}, the numerical coefficients of these monomials are not finite degree rational functions 
of the $p_i$. Indeed, already the simplest triple integrals in \eqn{B000} includes harmonic numbers, so we 
expect the coefficients to be expressible in terms of harmonic numbers, but we could not guess a simple form.

Another difference from \eqref{ep4asnew} is that these coefficients are actually not rational, they 
include both rational pieces and $\pi^{-2}$ times rational numbers. It is easy to see from the 
expressions in Appendix~ \ref{app:system}, that the triple integral in \eqn{gen_2} (with its prefactor) 
is proportional to $\pi^{-2}$, the square of the one-loop order is clearly rational, and the 
quadruple integral includes both rational and $\pi^{-2}$ pieces. In fact, the $\pi^0$ part of the quadruple 
integral is easy to trace, and is related in a simple way to the one-loop integral, as follows.

The integrand of the quadruple integral is the same as the square of the one-loop integrand, 
and as mentioned above equation \eqn{def_A}, once one chooses a particular set of Fourier 
modes, the numerator factorizes to cancel the denominator, leading to the integral \eqn{D00}. 
The resulting expression there seems rational, and further divided by 
$64\pi^4$, but positive powers of $\pi$ appear from degenerations in the coefficients $n_i$ 
(and it vanishes without those degenerations). A factor of $\pi^4$ comes only from full 
degeneration $n_i\to0$, giving the measure factor $\pi^4/24$ (which is divided by $64\pi^4$). 
Since the one-loop answer is solely from the zero-mode in the double integral and the 
integrands are the same, we find that the $\pi^0$ part of the last line in \eqn{gen_2} is equal 
to $-(\vev{W}_\lambda)^2/6$.

Combining with the square of the one-loop integral on the second line of \eqn{gen_2}, we get the 
very simple expression for the rational part of the two loop integral
\beq
\vev{W}_{\lambda^2}\Big|_{\pi^0} = \frac{1}{3}\left(\vev{W}_\lambda\right)^2\,.
\eeq
Note that this statement is true to all orders in $\epsilon$, so the difficulty is in calculating the 
$\pi^{-2}$ part in the two-loop VEV. Clearly the $\epsilon^2$ part in \eqn{2brem}, which is 
just the two-loop Bremsstrahlung function, is rational and is given by this recursion. 
Likewise, we expect at $k$-th loop order to find terms proportional to $\pi^0,\cdots , \pi^{2-2k}$, 
and all but the last term would be related to iterations of the lower loop expressions.


\section{Spectral dependence at higher orders in $\epsilon$}
\label{sec:ep6}

In the cases studied in \cite{Dekel:2015bla} there was spectral dependence starting from order $\epsilon^8$. 
As shown in the previous section, there is no dependence at order $\epsilon^4$, so here we address the 
possible spectral dependence at order $\epsilon^6$ and further patterns of spectral dependence at higher 
orders.

It should be possible to repeat the full analysis in the previous section for order $\epsilon^6$ with a bit more 
computing power, but we chose to do a partial survey by considering binomial $f(z)$, as in (\ref{fbinomial}).
Examining the cases with $2\leq q<p\leq22$ at one-loop order, we found that there was indeed spectral 
dependence in some examples at order $\epsilon^6$. The only such dependence is of the form 
$e^{\pm 2i\varphi}$ and happens only for $p=2q$ or $p=3q$. Specifically, the answer is a linear 
combination of terms of the form
\bal
\label{ep6dep}
&|a_{q}|^6\,,
&&
|a_{p}|^2|a_{q}|^4\,,
&&
|a_{p}|^4|a_{q}|^2\,,
\\
&\bar a_{2q}^2 a_q^4 e^{2i\varphi}\,,
&&
\bar a_{3q}|a_{3q}|^2 a_q^3 e^{2i\varphi}\,,
&&
\bar a_{3q}a_q^3|a_q|^2 e^{2i\varphi}\,,
\\
&a_{2q}^2\bar a_q^4 e^{-2i\varphi}\,,
\quad&&
a_{3q}|a_{3q}|^2 \bar a_q^3 e^{-2i\varphi}\,,
\quad&&
a_{3q}\bar a_q^3|a_q|^2 e^{-2i\varphi}\,.
\eal
At one-loop the coefficients are rational functions of $q$ and $p$, a bit more complicated than in \eqn{ep4asnew}, 
and at two-loops (where we did a far less extensive survey) they are harmonic numbers.

It is obvious that $e^{2i\varphi}$ would appear in terms where the number of $a$'s is two more than 
$\bar a$. But we see that there are no terms with $e^{4i\varphi}$, which would require five $a$'s and one 
$\bar a$. Furthermore we see that in all cases the total sum of $p$ and $q$ (weighted by degeneracy 
and a negative sign for $\bar a$) vanishes. We find then a pattern that generalizes the result at order 
$\epsilon^4$ and could hold for general order.
At order $\epsilon^{2l}$ we therefore expect only terms of the form
\beq
\label{conj}
a_{p_1}\cdots a_{p_j}\bar a_{q_1}\cdots\bar a_{q_{2l-j}}e^{2(j-l)i\varphi}\,,
\eeq
with the constraints
\beq
2\leq j\leq 2l-2\,,
\qquad
\sum_{i=1}^j p_i-\sum_{i=1}^{2l-j}q_i = 0\,.
\eeq
Clearly for $l=2$, one has $j=2$ and no spectral dependence. For $l=3$ there are many solutions, but if we 
restrict to binomials, so $p_i$ and $q_i$ can take only two values, we find indeed only the solutions in 
\eqn{ep6dep}.

In the binomial examples we checked beyond this order and at order $\epsilon^8$, 
we have found terms consistent with, among others,
\beq
\bar a_{3q}^2 a_q^6 e^{4i\varphi}\,,
\qquad
\bar a_{5q} a_q^5 |a_q|^2e^{4i\varphi}\,,
\qquad
\bar a_{5q}^3 a_{3q}^5 e^{2i\varphi}\,.
\eeq

\section{Strong coupling} 
\label{sec:strong}

One can also study perturbations of the circular Wilson loop at strong coupling, where 
it is described by a classical string in $AdS_3$. The VEV of the Wilson loop is given by 
the regularized area of the string, which is expressed in terms of $f(z)$ and 
$\alpha(z,\bar z)$ in \eqn{area}. As mentioned before, this expression is independent 
of the spectral parameter $\varphi$ to any order in $\epsilon$, but given the expressions 
above, we can plug in the expressions for $f(z)$ in \eqref{fofz} and 
$\alpha^{(2)}(z,\bar{z})$ in \eqref{alpha_hom_inhom} to find the VEV in 
terms of $a_p$'s.

This was done for a particular deformation of the circle 
in \cite{Huang:2016atz}, based on the perturbative algorithm of 
\cite{Dekel:2015bla} reviewed above. Here this is repeated for arbitrary $a_p$ 
to quartic order in $\epsilon$, giving
\bal
\label{string}
A_\text{reg}&=-2\pi\sqrt\lambda
-8\pi\sqrt\lambda\epsilon^2
\sum _{p=2}^{\infty }\frac{|a_{p}|^2}{p (p^2-1)}
\\&\quad
-64\pi\sqrt\lambda\epsilon^4\sum _{p\geq q\geq r\geq2}S^{(2)}_{p,q,r}
\frac{a_{p} \bar a_{q}\bar{a}_{r} a_{q+r-p}
+ \bar{a}_{p} a_{q} a_{r}\bar a_{q+r-p}}
{p(p^2-1)q(q^2-1)r(r^2-1)(q+r) ((q+r)^2-1)}
\\&\hskip3cm \times
\Big[p^2(q+r) ((q+r)^2-1)+p q r(q^2+3 q r+r^2+1)
\\&\hskip3.8cm
-(q+r)(q^4+q^3 r-q^2 r^2+q r^3+r^4-q^2-qr-r^2)\Big]\,,
\eal
with $S^{(2)}_{p,q,r}$ defined in \eqn{Snml}.
Note the similarities to \eqn{ep4asnew}, with the same denominators and the only difference being 
the polynomial in $p$, $q$ and $r$ in the numerator. In both expressions this is a degree 5 polynomial 
symmetric in $(q,r)$ with $p$ of at most degree 2.


\section{Discussion}
\label{sec:discussion}

In this paper we undertook to analyze the expectation values of deformed circular Wilson loops in 
$\cN=4$ SYM to quartic order in the deformation parameter. We presented explicit expressions 
for the expectation values at one-loop at weak coupling and at leading order at strong coupling. 
At two-loop order we provided all details of an explicit algorithm that calculates the VEV for 
particular deformations very efficiently.

As the stated goal of this project we have confirmed that the expectation value is independent 
of the spectral parameter at this order in the deformation and the perturbative orders that we 
examined. We expect this to hold to all orders in perturbation theory.

At higher orders in the deformation ($\epsilon^6$ and above), we found that generically 
there is spectral dependence in the 
expectation value of the Wilson loop, so different Wilson loop that are related to each other via 
this deformation, an exact strong-coupling symmetry, do not share a VEV at weak coupling.

Still, the pattern of spectral dependence as expressed in equation \eqn{conj}, which is the 
pattern we conjecture to hold to any order in the deformation, is exceedingly restrictive and simple. 
It is a striking result of our examinations that when representing the Wilson loop in terms of the 
sigma-model (Pohlmeyer) representation, \emph{i.e.}, in terms of $e^{i\varphi}f(z)$, there is such 
a simple pattern of allowed terms in the expression for the VEV of the Wilson loop at weak coupling. 
Furthermore, the one-loop expressions at order $\epsilon^4$ and at strong coupling share a 
very similar functional form as rational functions of the orders of terms in the Taylor expansion 
of $f(z)$.

It is very natural that the result of integration requires the sum of the Fourier coefficients to 
vanish. This precludes direct interaction between one very high Fourier mode and several 
low ones, a natural structure limiting UV-IR mixing. It is more surprising that similar 
structures arise, at weak coupling, in the calculation in terms of the Taylor coefficients 
of $f(z)$. Their weighted sum also seems to vanish and the spectral parameter can appear 
only in very specific combinations. This again can be thought of as avoiding UV-IR mixing, 
though it is surprising that it survives from the string description to weak coupling.

The spectral independence at order $\epsilon^4$ and other structures are hints to hidden 
symmetries in the expectation values of Wilson loop operators, which presumably have to do with 
all-loop integrability. A formalism exists for the all-loop integrability of the anomaly of cusps in the 
Wilson loop and for dimensions of insertions into the Wilson loop, but not for the question at hand
\cite{drukker, Drukker:2012de, Correa:2012hh, Gromov:2012eu, Gromov:2015dfa}, 
related to smooth Wilson loops. Perhaps there is a way of recombining results for 
3-point (and higher) correlators of insertions into the Wilson loop that would prove our 
conjectures \cite{Liendo:2016ymz,Cooke:2017qgm, Giombi:2017cqn,
Cavaglia:2018lxi,Giombi:2018qox,Liendo:2018ukf, Giombi:2018hsx}.
It would also be very 
interesting to understand any possible connections to other discussions of all-loop 
Yangian symmetry of smooth Wilson loops, as in \cite{Klose:2016uur, Klose:2016qfv} 
as well as the TBA formulation of the problem in \cite{Toledo:2014koa}.

An obvious generalization of our calculation is to study Wilson loops not restricted to $\bR^2$. 
This can be done, since it is understood how to solve for the boundary curve for open strings 
in the full $AdS_5$ \cite{He:2017cwd}. Though there is no explicit solution of the string 
sigma-model, as in the $AdS_3$ case, there is enough information to describe the boundary 
curve from details of a classical integrable model.

The same $AdS_3$ sigma-model describes also Wilson loops in ABJM theory restricted to a plane, 
so the strong coupling analysis applies directly there as well (with the appropriate redefinition of the 
coupling). This should capture the strong coupling limit of both locally $1/6$ BPS 
\cite{Drukker:2008zx, Chen:2008bp, Rey:2008bh} and $1/2$ BPS \cite{Drukker:2009hy}
Wilson loops of this theory. It should be possible to repeat our two-loop analysis for these loops and 
see what is the spectral dependence in the two classes of loops and whether they are related 
to each other. 

It may be possible to extend our strong coupling analysis beyond the leading order and include the 
first string fluctuations, as was done for cusps in \cite{Chu:2009qt,Forini:2010ek,Drukker:2011za} and 
for circular Wilson loops in \cite{Drukker:2000ep, Forini:2015bgo, Faraggi:2016ekd, 
Forini:2017whz, Cagnazzo:2017sny, Medina-Rincon:2018wjs}. It would be 
very interesting to see whether spectral dependence (at order $\epsilon^6$) appears in that 
calculation.


\section*{Acknowledgments}

We acknowledge useful discussions with Bartomeu Fiol, Mart\'in Kruczenski, 
Georgios Papathanasiou and Kostya Zarembo. 
The work of A.D. was supported by the Swedish Research Council (VR) grant 2013-4329. 
The research of N.D. was supported in part by the Science \& Technology Facilities Council 
via the consolidated grant number ST/P000258/1 and by a joint KCL-FAPESP fellowship. He 
is grateful to the hospitality of the University of S\~ao Paulo, Peking University, Weizmann Institute, 
Nordita and IIP Natal during the course of this work.
D.T. acknowledges partial financial support from CNPq, the FAPESP grants 
2014/18634-9 and 2015/17885-0,  the joint KCL-FAPESP grant 2017/50435-4, and the FAPESP grant 2016/01343-7 through ICTP-SAIFR. 
The work of E.V. is funded by the FAPESP grants 2014/18634-9 and 2016/09266-1, and
by the STFC grant ST/P000762/1. 
He thanks Yunfeng Jiang and ETH Zurich for hospitality while this project was in preparation.

\appendix

\section{Two-loop integrals}
\label{app:system}

In this appendix we present the details of the two-loop calculation, which amounts to finding the Fourier 
representation of \eqn{gen_2}. Expanding the integrands in powers of $g(\theta)$ 
(see \eqn{fourier}) gives several types of basic integrals studied below.

\subsection{Logarithmic integrals}
\label{app:system_2}

The first set of integrals encountered in the expansion of \eqn{gen_2} in powers 
of $b_{n_i}$ are
\beq
\label{def_B}
B_{n_{1},n_{2},n_{3}}^{p_1,p_2,p_3}
=-\frac{i}{8\pi^2}
\oint d\theta_1 d\theta_2 d\theta_3 \,\varepsilon(\theta_{1},\theta_{2},\theta_{3})
\frac{e^{i(n_{1}\theta_{1}+n_{2}\theta_{2}+n_{3}\theta_{3})}
\log\left(2-2\cos(\theta_{1}-\theta_{2})\right)}
{(e^{i\theta_{1}}-e^{i\theta_{2}})^{p_1}(e^{i\theta_{2}}-e^{i\theta_{3}})^{p_2}
(e^{i\theta_{3}}-e^{i\theta_{1}})^{p_3}}\,.
\eeq
For vanishing $p$'s, the integrals $B_{n_{1},n_{2},n_{3}}^{0,0,0}$ are quite simple. 
We perform the calculation for real $n_i$ and take the limit to the integers at the end.

Expanding the antisymmetric symbol and integrating over $\theta_3$ gives
\bal
& \frac{1}{8\pi^2n_{3}}\int_0^{2\pi} d\theta_1 \int_0^{\theta_1} 
d\theta_2 \left(e^{i(n_{1}\theta_{1}+n_{2}\theta_{2})}-e^{i(n_{2}\theta_{1}+n_{1}\theta_{2})}\right)
\left(2e^{in_{3}\theta_{1}}-2e^{in_{3}\theta_{2}}+1-e^{2\pi in_{3}}\right)\\
&\quad\times
\log\left(2-2\cos\left(\theta_{1}-\theta_{2}\right)\right).
\eal
The change of variables $\theta_{1}=(u-v)/2,\,\theta_{2}=(u+v)/2$ allows to integrate 
over $u$, resulting in
\bal
\label{label1}
-\frac{i}{8\pi^2} \int_{-2\pi}^{0}dv
\bigg(&\frac{2(e^{in_{3}v}-1)(e^{2\pi i(n_{1}+n_{2}+n_{3})}
-e^{-i(n_{1}+n_{2}+n_{3})v})}{n_{1}+n_{2}+n_{3}}
\\&+\frac{(e^{2\pi in_{3}}-1)(e^{2\pi i(n_{1}+n_{2})}-e^{-i(n_{1}+n_{2})v})}{n_{1}+n_{2}}\bigg)
\frac{e^{in_{1}v}-e^{in_{2}v}}{n_{3}}\log(2-2\cos v)\,. 
\eal
Now we can take the $n_i$ to be integer, noting that no poles arise even when the sums of $n_i$ in 
the denominators vanish. The result of the $v$ integration is then
\begingroup \allowdisplaybreaks 
\begin{gather}\label{B000}
B_{n_{1},n_{2},n_{3}}^{0,0,0} =
\begin{cases}
\displaystyle
\frac{(|n_{1}|-|n_{2}|)}{|n_{1}n_{2}|(n_{1}+n_{2})} \,,
& \quad n_{1}+n_{2}+n_3=0, \quad n_{1}+n_{2}\neq0, 
\quad n_1\neq 0, \quad n_2\neq 0\,,
\hskip-3mm
\\[4mm]
\displaystyle
-\frac{2H_{|n_{1}|-1}}{n_{1}}\,, & \quad n_{1}+n_{2}=n_3=0, \quad n_1\neq 0\,,\\
\displaystyle
-\sign(n_{1})\frac{1}{n_{1}^{2}}\,, & \quad n_1+n_3=n_2=0, \quad n_1\neq0\,,\\
\displaystyle
\sign(n_{2})\frac{1}{n_{2}^{2}}\,, & \quad n_2+n_3=n_1=0, \quad n_2\neq0\,,\\
0\,, & \quad \text{otherwise.}
\end{cases}
\end{gather}
\endgroup
Here $H_n = \sum_{m=1}^{\infty} 1/m$ are the harmonic numbers.

For positive $p_1,p_2,p_3$, we do not evaluate $B_{n_{1},n_{2},n_{3}}^{p_1,p_2,p_3}$ by 
integration, but rather impose the consistency and recursion relations
\begin{gather}\label{eq_for_B}
B_{n_{1},n_{2},n_{3}}^{p_1,p_2,p_3}=(-1)^{p_1+p_2+p_3+1}B_{n_{2},n_{1},n_{3}}^{p_1,p_3,p_2}\,,\\
\sum_{k_{1}=0}^{p_{1}}\sum_{k_{2}=0}^{p_{2}}\sum_{k_{3}=0}^{p_{3}}
\binom{p_1}{k_1}
\binom{p_2}{k_2}
\binom{p_3}{k_3}
(-1)^{k_{1}+k_{2}+k_{3}}
B_{n_{1}+p_{1}+k_{3}-k_{1},n_{2}+p_{2}+k_{1}-k_{2},n_{3}+p_{3}+k_{2}-k_{3}}^{p_{1},p_{2},p_{3}}
= B_{n_{1},n_{2},n_{3}}^{0,0,0}\,.\nonumber 
\end{gather}

The coefficients appearing in the calculation of the Wilson loop to quadratic order in 
$b_{n_i}$ are $B_{n_1,n_2,n_3}^{1,2,2}$. Restricting by parity symmetry to 
$n_{1}\leq n_{2}$ we find that:
\begin{itemize}
\item
For $n_1\leq n_{2}\leq n_{1}+3$ one can choose the coefficients $B_{n_1,n_2,n_3}^{1,2,2}$ 
freely, and we set them to zero for simplicity.

\item
The coefficients with $n_{2}>n_{1}+3$ are generated recursively through equation 
\eqref{eq_for_B} (with the replacements $n_{2}\to n_{2}-3$, $n_{3}\to n_{3}-2$)
\begingroup \allowdisplaybreaks 
\bal
\label{label2}
B_{n_{1},n_{2},n_{3}}^{1,2,2} & =-B_{n_{1},n_{2}-3,n_{3}-2}^{0,0,0}
-B_{n_{1},n_{2}-2,n_{3}+2}^{1,2,2}+2B_{n_{1},n_{2}-1,n_{3}+1}^{1,2,2}
+B_{n_{1}+1,n_{2}-3,n_{3}+2}^{1,2,2}\\
&\quad{}
-3B_{n_{1}+1,n_{2}-1,n_{3}}^{1,2,2}+2B_{n_{1}+1,n_{2},n_{3}-1}^{1,2,2}-2B_{n_{1}
+2,n_{2}-3,n_{3}+1}^{1,2,2}+3B_{n_{1}+2,n_{2}-2,n_{3}}^{1,2,2}
\\
&\quad{}
-B_{n_{1}+2,n_{2},n_{3}-2}^{1,2,2}+B_{n_{1}+3,n_{2}-3,n_{3}}^{1,2,2}-2B_{n_{1}+3,n_{2}-2,n_{3}-1}^{1,2,2}+B_{n_{1}+3,n_{2}-1,n_{3}-2}^{1,2,2}\,.
\eal
\endgroup
\end{itemize}

At quartic order in the $b_{n_i}$ we need the coefficients $B_{n_1,n_2,n_3}^{3,4,4}$. 
Taking, without loss in generality, $n_{1}\leq n_{2}$ we have
\begin{itemize}
\item
The coefficients with $n_1\leq n_{2}\leq n_{1}+7$ can be chosen freely and we set them to zero.

\item
For $n_{1}+7<n_{2}$ the coefficients are generated recursively from \eqref{eq_for_B}:
\end{itemize}
\begingroup \allowdisplaybreaks 
\bal\label{label10}
B_{n_{1},n_{2},n_{3}}^{3,4,4} & =-B_{n_{1},n_{2}-7,n_{3}-4}^{0,0,0}
-B_{n_{1},n_{2}-4,n_{3}+4}^{3,4,4}+4B_{n_{1},n_{2}-3,n_{3}+3}^{3,4,4}
-6B_{n_{1},n_{2}-2,n_{3}+2}^{3,4,4}
\\&\quad
+4B_{n_{1},n_{2}-1,n_{3}+1}^{3,4,4}+3B_{n_{1}+1,n_{2}-5,n_{3}+4}^{3,4,4}
-8B_{n_{1}+1,n_{2}-4,n_{3}+3}^{3,4,4}+2B_{n_{1}+1,n_{2}-3,n_{3}+2}^{3,4,4}
\\&\quad
+12B_{n_{1}+1,n_{2}-2,n_{3}+1}^{3,4,4}-13B_{n_{1}+1,n_{2}-1,n_{3}}^{3,4,4}
+4B_{n_{1}+1,n_{2},n_{3}-1}^{3,4,4}-3B_{n_{1}+2,n_{2}-6,n_{3}+4}^{3,4,4}
\\&\quad
+24B_{n_{1}+2,n_{2}-4,n_{3}+2}^{3,4,4}-36B_{n_{1}+2,n_{2}-3,n_{3}+1}^{3,4,4}
+9B_{n_{1}+2,n_{2}-2,n_{3}}^{3,4,4}+12B_{n_{1}+2,n_{2}-1,n_{3}-1}^{3,4,4}
\!\!\!
\\&\quad
-6B_{n_{1}+2,n_{2},n_{3}-2}^{3,4,4}+B_{n_{1}+3,n_{2}-7,n_{3}+4}^{3,4,4}
+8B_{n_{1}+3,n_{2}-6,n_{3}+3}^{3,4,4}-24B_{n_{1}+3,n_{2}-5,n_{3}+2}^{3,4,4}
\\&\quad
+45B_{n_{1}+3,n_{2}-3,n_{3}}^{3,4,4}-36B_{n_{1}+3,n_{2}-2,n_{3}-1}^{3,4,4}
+2B_{n_{1}+3,n_{2}-1,n_{3}-2}^{3,4,4}+4B_{n_{1}+3,n_{2},n_{3}-3}^{3,4,4}
\\&\quad
-4B_{n_{1}+4,n_{2}-7,n_{3}+3}^{3,4,4}-2B_{n_{1}+4,n_{2}-6,n_{3}+2}^{3,4,4}
+36B_{n_{1}+4,n_{2}-5,n_{3}+1}^{3,4,4}-45B_{n_{1}+4,n_{2}-4,n_{3}}^{3,4,4}
\\&\quad
+24B_{n_{1}+4,n_{2}-2,n_{3}-2}^{3,4,4}-8B_{n_{1}+4,n_{2}-1,n_{3}-3}^{3,4,4}
-B_{n_{1}+4,n_{2},n_{3}-4}^{3,4,4}+6B_{n_{1}+5,n_{2}-7,n_{3}+2}^{3,4,4}
\\&\quad
-12B_{n_{1}+5,n_{2}-6,n_{3}+1}^{3,4,4}-9B_{n_{1}+5,n_{2}-5,n_{3}}^{3,4,4}
+36B_{n_{1}+5,n_{2}-4,n_{3}-1}^{3,4,4}-24B_{n_{1}+5,n_{2}-3,n_{3}-2}^{3,4,4}
\!\!\!
\\&\quad
+3B_{n_{1}+5,n_{2}-1,n_{3}-4}^{3,4,4}-4B_{n_{1}+6,n_{2}-7,n_{3}+1}^{3,4,4}
+13B_{n_{1}+6,n_{2}-6,n_{3}}^{3,4,4}-12B_{n_{1}+6,n_{2}-5,n_{3}-1}^{3,4,4}
\\&\quad
-2B_{n_{1}+6,n_{2}-4,n_{3}-2}^{3,4,4}+8B_{n_{1}+6,n_{2}-3,n_{3}-3}^{3,4,4}
-3B_{n_{1}+6,n_{2}-2,n_{3}-4}^{3,4,4}+B_{n_{1}+7,n_{2}-7,n_{3}}^{3,4,4}
\\&\quad
-4B_{n_{1}+7,n_{2}-6,n_{3}-1}^{3,4,4}+6B_{n_{1}+7,n_{2}-5,n_{3}-2}^{3,4,4}
-4B_{n_{1}+7,n_{2}-4,n_{3}-3}^{3,4,4}+B_{n_{1}+7,n_{2}-3,n_{3}-4}^{3,4,4}\,. 
\eal
\endgroup

\subsection{Non-logarithmic part in two-loop interacting diagrams}
\label{app:system_3}
Another class of integrals arising from the expansion of the first term in 
\eqn{gen_2} in powers of the Fourier coefficients $b_{n_i}$ are
\beq
\label{def_C}
C_{n_{1},n_{2},n_{3}}^{p_{1},p_{2},p_{3}}=
-\frac{i}{8\pi^2}
\oint d\theta_{1} d\theta_2 d\theta_3 \, \varepsilon(\theta_{1},\theta_{2},\theta_{3})
\frac{e^{in_{1}\theta_{1}+in_{2}\theta_{2}+in_{3}\theta_{3}}}
{(e^{i\theta_{1}}-e^{i\theta_{2}})^{p_{1}}(e^{i\theta_{2}}-e^{i\theta_{3}})^{p_{2}}
(e^{i\theta_{3}}-e^{i\theta_{1}})^{p_{3}}}\,.
\eeq
The integral with all $p$'s vanishing is finite and evaluates to
\begingroup \allowdisplaybreaks 
\begin{gather} \label{C000}
C_{n_{1},n_{2},n_{3}}^{0,0,0}=
\begin{cases}
\displaystyle
\frac{1}{n_2} & \quad n_2+n_3=n_1=0, \quad n_2\neq0\vspace{0.2 cm}\\
\displaystyle
\frac{1}{n_3} & \quad n_1+n_3=n_2=0, \quad n_3\neq0\vspace{0.2 cm}\\
\displaystyle
\frac{1}{n_1} & \quad n_1+n_2=n_3=0, \quad n_1\neq0\\
0 & \quad \text{otherwise.}
\end{cases}
\end{gather}
\endgroup

The other integrals are divergent, but as explained before, we expect the divergences to cancel in the final expressions 
and we choose a ``regularization'' of them by requiring the symmetry and recursion relations
\beq
\begin{gathered}
C_{n_{1},n_{2},n_{3}}^{p_{1},p_{2},p_{3}}
=(-1)^{p_{1}+p_{2}+p_{3}+1}C_{n_{2},n_{1},n_{3}}^{p_{1},p_{3},p_{2}}
=(-1)^{p_{1}+p_{2}+p_{3}+1}C_{n_{1},n_{3},n_{2}}^{p_{3},p_{2},p_{1}}
=(-1)^{p_{1}+p_{2}+p_{3}+1}C_{n_{3},n_{2},n_{1}}^{p_{2},p_{1},p_{3}}\,,
\\
\sum_{k_{1}=0}^{p_{1}}\sum_{k_{2}=0}^{p_{2}}\sum_{k_{3}=0}^{p_{3}}
\binom{p_1}{k_1}
\binom{p_2}{k_2}
\binom{p_3}{k_3}
(-1)^{k_{1}+k_{2}+k_{3}}
C_{n_{1}+p_{1}+k_{3}-k_{1},n_{2}+p_{2}+k_{1}-k_{2},n_{3}+p_{3}+k_{2}-k_{3}}^{p_{1},p_{2},p_{3}} 
= C_{n_{1},n_{2},n_{3}}^{0,0,0}\,.
\label{eq_for_C}
\end{gathered}
\eeq

The relevant building blocks for the quartic order in $b_{n_i}$ are 
$C_{n_{1},n_{2},n_{3}}^{3,3,3}$. We restrict to $n_{1}\leq n_{2}\leq n_{3}$ 
thanks to the first line in \eqref{eq_for_C}.
\begin{itemize}
\item
The coefficients with $n_{3}-3\leq n_{2}$ and with $n_{2}-3\leq n_{1}$ are undetermined by the recursion 
relations and we can set them to zero for simplicity.
\item
The coefficients with $n_{3}> n_{2}+3> n_{1}+6$ are generated through 
\eqref{eq_for_C} after shifting $n_{2}\to n_{2}-3,\,n_{3}\to n_{3}-6$
\end{itemize}
\begingroup \allowdisplaybreaks 
\bal\label{label11}
C_{n_{1},n_{2},n_{3}}^{3,3,3} & =C_{n_{1},n_{2}-3,n_{3}-6}^{0,0,0}
+3C_{n_{1},n_{2}+1,n_{3}-1}^{3,3,3}-3C_{n_{1},n_{2}+2,n_{3}-2}^{3,3,3}
+C_{n_{1},n_{2}+3,n_{3}-3}^{3,3,3}
\\&\quad
+3C_{n_{1}+1,n_{2}-1,n_{3}}^{3,3,3}-6C_{n_{1}+1,n_{2},n_{3}-1}^{3,3,3}
+6C_{n_{1}+1,n_{2}+2,n_{3}-3}^{3,3,3}-3C_{n_{1}+1,n_{2}+3,n_{3}-4}^{3,3,3} 
\\&\quad
-3C_{n_{1}+2,n_{2}-2,n_{3}}^{3,3,3}+15C_{n_{1}+2,n_{2},n_{3}-2}^{3,3,3}
-15C_{n_{1}+2,n_{2}+1,n_{3}-3}^{3,3,3}+3C_{n_{1}+2,n_{2}+3,n_{3}-5}^{3,3,3} 
\\&\quad
+C_{n_{1}+3,n_{2}-3,n_{3}}^{3,3,3}+6C_{n_{1}+3,n_{2}-2,n_{3}-1}^{3,3,3}
-15C_{n_{1}+3,n_{2}-1,n_{3}-2}^{3,3,3}+15C_{n_{1}+3,n_{2}+1,n_{3}-4}^{3,3,3} 
\\&\quad
-6C_{n_{1}+3,n_{2}+2,n_{3}-5}^{3,3,3}-C_{n_{1}+3,n_{2}+3,n_{3}-6}^{3,3,3}
-3C_{n_{1}+4,n_{2}-3,n_{3}-1}^{3,3,3}+15C_{n_{1}+4,n_{2}-1,n_{3}-3}^{3,3,3} 
\\&\quad
-15C_{n_{1}+4,n_{2},n_{3}-4}^{3,3,3}+3C_{n_{1}+4,n_{2}+2,n_{3}-6}^{3,3,3}
+3C_{n_{1}+5,n_{2}-3,n_{3}-2}^{3,3,3}-6C_{n_{1}+5,n_{2}-2,n_{3}-3}^{3,3,3} 
\\&\quad
+6C_{n_{1}+5,n_{2},n_{3}-5}^{3,3,3}-3C_{n_{1}+5,n_{2}+1,n_{3}-6}^{3,3,3}
-C_{n_{1}+6,n_{2}-3,n_{3}-3}^{3,3,3}+3C_{n_{1}+6,n_{2}-2,n_{3}-4}^{3,3,3} 
\\&\quad
-3C_{n_{1}+6,n_{2}-1,n_{3}-5}^{3,3,3}+C_{n_{1}+6,n_{2},n_{3}-6}^{3,3,3}\,. 
\eal
\endgroup

\subsection{Two-loop ladder diagrams}
\label{app:system_4}
The quadruple ordered integral in the ladder contribution to \eqref{gen_2} decomposes into a sum of
\beq
\label{def_D}
D_{n_{1},n_{2},n_{3},n_{4}}^{p_{1},p_{2}}
=
\int_0^{2\pi} d\theta_1 \int_0^{\theta_1} d\theta_2 \int_0^{\theta_2} d\theta_3 
\int_0^{\theta_3} d\theta_4\,
\frac{e^{i(n_{1}\theta_{1}+n_{2}\theta_{2}+n_{3}\theta_{3}+n_{4}\theta_{4})}}
{(e^{i\theta_{1}}-e^{i\theta_{3}})^{p_{1}}(e^{i\theta_{2}}-e^{i\theta_{4}})^{p_{2}}}\,.
\eeq
The simplest integral of the family
\begingroup \allowdisplaybreaks 
\begin{flalign}
\label{D00}
D_{n_{1},n_{2},n_{3},n_{4}}^{0,0} 
& =\frac{1}{n_{1}\left(n_{1}+n_{2}\right)\left(n_{1}+n_{2}+n_{3}\right)\left(n_{1}+n_{2}+n_{3}+n_{4}\right)}-\frac{e^{2\pi in_{1}}}{n_{1}n_{2}\left(n_{2}+n_{3}\right)\left(n_{2}+n_{3}+n_{4}\right)}\nonumber 
\\&\quad
+\frac{e^{2\pi i\left(n_{1}+n_{2}\right)}}{n_{2}n_{3}\left(n_{1}+n_{2}\right)\left(n_{3}+n_{4}\right)}-\frac{e^{2\pi i\left(n_{1}+n_{2}+n_{3}\right)}}{n_{3}n_{4}\left(n_{2}+n_{3}\right)\left(n_{1}+n_{2}+n_{3}\right)}\nonumber 
\\&\quad
+\frac{e^{2\pi i\left(n_{1}+n_{2}+n_{3}+n_{4}\right)}}{n_{4}\left(n_{3}+n_{4}\right)\left(n_{2}+n_{3}+n_{4}\right)\left(n_{1}+n_{2}+n_{3}+n_{4}\right)}
\end{flalign}
\endgroup
is non-singular for any integer $n_i$, with the appropriate application of L'H\^opital's rule, as before. 
We define the others through the relations
\begin{gather}\label{eq_for_D}
\sum_{k_{1}=0}^{p_{1}}\sum_{k_{2}=0}^{p_{2}}
\binom{p_1}{k_1}
\binom{p_2}{k_2}
(-1)^{k_{1}+k_{2}}
D_{n_{1}+p_{1}-k_{1},n_{2}+p_{2}-k_{2},n_{3}+k_{1},n_{4}+k_{2}}^{p_{1},p_{2}}
= D_{n_{1},n_{2},n_{3},n_{4}}^{0,0}\,. 
\end{gather}
The ordered integrations \eqref{def_D} imply no parity for the integrals, which means that more 
recurrence formulas are needed to cover all values of $n_i$. We start with the solution for 
$D_{n_{1},n_{2},n_{3},n_{4}}^{2,2}$.
\begin{itemize}
\item
The coefficients with $0\leq n_{3}-n_{1}\leq3$ or $0\leq n_{2}\leq1$ are arbitrary constants.
\item
The coefficients with $4\leq n_{3}-n_{1}$ and $2\leq n_{2}$ are given by
\begingroup \allowdisplaybreaks 
\bal
\label{label3}
D_{n_{1},n_{2},n_{3},n_{4}}^{2,2} & =D_{n_{1},n_{2}-2,n_{3}-2,n_{4}}^{0,0}-D_{n_{1},n_{2}-2,n_{3},n_{4}+2}^{2,2}+2D_{n_{1},n_{2}-1,n_{3},n_{4}+1}^{2,2}
\\&\quad
+2D_{n_{1}+1,n_{2}-2,n_{3}-1,n_{4}+2}^{2,2}-4D_{n_{1}+1,n_{2}-1,n_{3}-1,n_{4}+1}^{2,2}+2D_{n_{1}+1,n_{2},n_{3}-1,n_{4}}^{2,2} 
\\&\quad
-D_{n_{1}+2,n_{2}-2,n_{3}-2,n_{4}+2}^{2,2}+2D_{n_{1}+2,n_{2}-1,n_{3}-2,n_{4}+1}^{2,2}-D_{n_{1}+2,n_{2},n_{3}-2,n_{4}}^{2,2}\,. 
\eal
\endgroup
\item
For $n_{3}-n_{1}\leq-1$ and $2\leq n_{2}$ we find
\begingroup \allowdisplaybreaks 
\bal
D_{n_{1},n_{2},n_{3},n_{4}}^{2,2} & =D_{n_{1}-2,n_{2}-2,n_{3},n_{4}}^{0,0}
-D_{n_{1}-2,n_{2}-2,n_{3}+2,n_{4}+2}^{2,2}+2D_{n_{1}-2,n_{2}-1,n_{3}+2,n_{4}+1}^{2,2}
\\&\quad
-D_{n_{1}-2,n_{2},n_{3}+2,n_{4}}^{2,2}+2D_{n_{1}-1,n_{2}-2,n_{3}+1,n_{4}+2}^{2,2}
-4D_{n_{1}-1,n_{2}-1,n_{3}+1,n_{4}+1}^{2,2} 
\\&\quad
+2D_{n_{1}-1,n_{2},n_{3}+1,n_{4}}^{2,2}-D_{n_{1},n_{2}-2,n_{3},n_{4}+2}^{2,2}
+2D_{n_{1},n_{2}-1,n_{3},n_{4}+1}^{2,2}\,. 
\eal
\endgroup
\item
For $n_{3}-n_{1}\leq-1$ and $n_{2}\leq-1$ we find 
\begingroup \allowdisplaybreaks 
\bal
D_{n_{1},n_{2},n_{3},n_{4}}^{2,2} & =D_{n_{1}-2,n_{2},n_{3},n_{4}-2}^{0,0}
-D_{n_{1}-2,n_{2},n_{3}+2,n_{4}}^{2,2}+2D_{n_{1}-2,n_{2}+1,n_{3}+2,n_{4}-1}^{2,2}
\\&\quad
-D_{n_{1}-2,n_{2}+2,n_{3}+2,n_{4}-2}^{2,2}+2D_{n_{1}-1,n_{2},n_{3}+1,n_{4}}^{2,2}
-4D_{n_{1}-1,n_{2}+1,n_{3}+1,n_{4}-1}^{2,2} 
\\&\quad
+2D_{n_{1}-1,n_{2}+2,n_{3}+1,n_{4}-2}^{2,2}+2D_{n_{1},n_{2}+1,n_{3},n_{4}-1}^{2,2}
-D_{n_{1},n_{2}+2,n_{3},n_{4}-2}^{2,2}\,. 
\eal
\endgroup
\item
Finally, the coefficients with $4\leq n_{3}-n_{1}$ and $n_{2}\leq-1$ are given by
\begingroup \allowdisplaybreaks 
\bal\label{label4}
D_{n_{1},n_{2},n_{3},n_{4}}^{2,2} & =D_{n_{1},n_{2},n_{3}-2,n_{4}-2}^{0,0}
+2D_{n_{1},n_{2}+1,n_{3},n_{4}-1}^{2,2}-D_{n_{1},n_{2}+2,n_{3},n_{4}-2}^{2,2}
\\&\quad
+2D_{n_{1}+1,n_{2},n_{3}-1,n_{4}}^{2,2}-4D_{n_{1}+1,n_{2}+1,n_{3}-1,n_{4}-1}^{2,2}
+2D_{n_{1}+1,n_{2}+2,n_{3}-1,n_{4}-2}^{2,2} 
\\&\quad
-D_{n_{1}+2,n_{2},n_{3}-2,n_{4}}^{2,2}+2D_{n_{1}+2,n_{2}+1,n_{3}-2,n_{4}-1}^{2,2}
-D_{n_{1}+2,n_{2}+2,n_{3}-2,n_{4}-2}^{2,2}\,. 
\eal
\endgroup
\end{itemize}
The order $\lambda^2\epsilon^4$ depends also on the elementary integrals $D_{n_{1},n_{2},n_{3},n_{4}}^{4,4}$.
\begin{itemize}
\item
The coefficients with $0\leq n_{3}-n_{1}\leq7$ or $0\leq n_{2}\leq3$
are arbitrary constants.
\item
The coefficients with $8\leq n_{3}-n_{1}$ and $4\leq n_{2}$ are generated recursively by
\begingroup \allowdisplaybreaks 
\begin{align}\label{label8}
D_{n_{1},n_{2},n_{3},n_{4}}^{4,4} & =D_{n_{1},n_{2}-4,n_{3}-4,n_{4}}^{0,0}-D_{n_{1},n_{2}-4,n
_{3},n_{4}+4}^{4,4}+4D_{n_{1},n_{2}-3,n_{3},n_{4}+3}^{4,4}-6D_{n_{1},n_{2}-2,n_{3},n_{4}+2}^{4,4}\nonumber
\\&\quad
+4D_{n_{1},n_{2}-1,n_{3},n_{4}+1}^{4,4}+4D_{n_{1}+1,n_{2}-4,n_{3}-1,n_{4}+4}^{4,4}-16D_{n_{1}+1,n_{2}-3,n_{3}-1,n_{4}+3}^{4,4}\nonumber 
\\&\quad
+24D_{n_{1}+1,n_{2}-2,n_{3}-1,n_{4}+2}^{4,4}-16D_{n_{1}+1,n_{2}-1,n_{3}-1,n_{4}+1}^{4,4}+4D_{n_{1}+1,n_{2},n_{3}-1,n_{4}}^{4,4}\nonumber 
\\&\quad
-6D_{n_{1}+2,n_{2}-4,n_{3}-2,n_{4}+4}^{4,4}+24D_{n_{1}+2,n_{2}-3,n_{3}-2,n_{4}+3}^{4,4}-36D_{n_{1}+2,n_{2}-2,n_{3}-2,n_{4}+2}^{4,4}\nonumber 
\\&\quad
+24D_{n_{1}+2,n_{2}-1,n_{3}-2,n_{4}+1}^{4,4}-6D_{n_{1}+2,n_{2},n_{3}-2,n_{4}}^{4,4}+4D_{n_{1}+3,n_{2}-4,n_{3}-3,n_{4}+4}^{4,4}\nonumber 
\\&\quad
-16D_{n_{1}+3,n_{2}-3,n_{3}-3,n_{4}+3}^{4,4}+24D_{n_{1}+3,n_{2}-2,n_{3}-3,n_{4}+2}^{4,4}-16D_{n_{1}+3,n_{2}-1,n_{3}-3,n_{4}+1}^{4,4}\nonumber 
\\&\quad
+4D_{n_{1}+3,n_{2},n_{3}-3,n_{4}}^{4,4}-D_{n_{1}+4,n_{2}-4,n_{3}-4,n_{4}+4}^{4,4}+4D_{n_{1}+4,n_{2}-3,n_{3}-4,n_{4}+3}^{4,4}\nonumber 
\\&\quad
-6D_{n_{1}+4,n_{2}-2,n_{3}-4,n_{4}+2}^{4,4}+4D_{n_{1}+4,n_{2}-1,n_{3}-4,n_{4}+1}^{4,4}-D_{n_{1}+4,n_{2},n_{3}-4,n_{4}}^{4,4}\,. 
\end{align}
\endgroup
\item
The coefficients with $n_{3}-n_{1}\leq-1$ and $4\leq n_{2}$ are generated recursively by
\begingroup \allowdisplaybreaks 
\begin{align}
D_{n_{1},n_{2},n_{3},n_{4}}^{4,4} & =D_{n_{1}-4,n_{2}-4,n_{3},n_{4}}^{0,0}-D_{n_{1}-4,n_{2}-4,n_{3}+4,n_{4}+4}^{4,4}+4D_{n_{1}-4,n_{2}-3,n_{3}+4,n_{4}+3}^{4,4}
\\&\quad
-6D_{n_{1}-4,n_{2}-2,n_{3}+4,n_{4}+2}^{4,4}+4D_{n_{1}-4,n_{2}-1,n_{3}+4,n_{4}+1}^{4,4}-D_{n_{1}-4,n_{2},n_{3}+4,n_{4}}^{4,4}\nonumber 
\\&\quad
+4D_{n_{1}-3,n_{2}-4,n_{3}+3,n_{4}+4}^{4,4}-16D_{n_{1}-3,n_{2}-3,n_{3}+3,n_{4}+3}^{4,4}+24D_{n_{1}-3,n_{2}-2,n_{3}+3,n_{4}+2}^{4,4}\nonumber 
\\&\quad
-16D_{n_{1}-3,n_{2}-1,n_{3}+3,n_{4}+1}^{4,4}+4D_{n_{1}-3,n_{2},n_{3}+3,n_{4}}^{4,4}-6D_{n_{1}-2,n_{2}-4,n_{3}+2,n_{4}+4}^{4,4}\nonumber 
\\&\quad
+24D_{n_{1}-2,n_{2}-3,n_{3}+2,n_{4}+3}^{4,4}-36D_{n_{1}-2,n_{2}-2,n_{3}+2,n_{4}+2}^{4,4}+24D_{n_{1}-2,n_{2}-1,n_{3}+2,n_{4}+1}^{4,4}\nonumber 
\\&\quad
-6D_{n_{1}-2,n_{2},n_{3}+2,n_{4}}^{4,4}+4D_{n_{1}-1,n_{2}-4,n_{3}+1,n_{4}+4}^{4,4}-16D_{n_{1}-1,n_{2}-3,n_{3}+1,n_{4}+3}^{4,4}\nonumber 
\\&\quad
+24D_{n_{1}-1,n_{2}-2,n_{3}+1,n_{4}+2}^{4,4}-16D_{n_{1}-1,n_{2}-1,n_{3}+1,n_{4}+1}^{4,4}+4D_{n_{1}-1,n_{2},n_{3}+1,n_{4}}^{4,4}\nonumber 
\\&\quad
-D_{n_{1},n_{2}-4,n_{3},n_{4}+4}^{4,4}+4D_{n_{1},n_{2}-3,n_{3},n_{4}+3}^{4,4}-6D_{n_{1},n_{2}-2,n_{3},n_{4}+2}^{4,4}+4D_{n_{1},n_{2}-1,n_{3},n_{4}+1}^{4,4}\,.\nonumber 
\end{align}
\endgroup
\item
The coefficients with $n_{3}-n_{1}\leq-1$ and $n_{2}\leq-1$ are generated recursively by
\begingroup \allowdisplaybreaks 
\begin{align}
D_{n_{1},n_{2},n_{3},n_{4}}^{4,4} & =D_{n_{1}-4,n_{2},n_{3},n_{4}-4}^{0,0}-D_{n_{1}-4,n_{2},n_{3}+4,n_{4}}^{4,4}+4D_{n_{1}-4,n_{2}+1,n_{3}+4,n_{4}-1}^{4,4}
\\&\quad
-6D_{n_{1}-4,n_{2}+2,n_{3}+4,n_{4}-2}^{4,4}+4D_{n_{1}-4,n_{2}+3,n_{3}+4,n_{4}-3}^{4,4}-D_{n_{1}-4,n_{2}+4,n_{3}+4,n_{4}-4}^{4,4}\nonumber 
\\&\quad
+4D_{n_{1}-3,n_{2},n_{3}+3,n_{4}}^{4,4}-16D_{n_{1}-3,n_{2}+1,n_{3}+3,n_{4}-1}^{4,4}+24D_{n_{1}-3,n_{2}+2,n_{3}+3,n_{4}-2}^{4,4}\nonumber 
\\&\quad
-16D_{n_{1}-3,n_{2}+3,n_{3}+3,n_{4}-3}^{4,4}+4D_{n_{1}-3,n_{2}+4,n_{3}+3,n_{4}-4}^{4,4}-6D_{n_{1}-2,n_{2},n_{3}+2,n_{4}}^{4,4}\nonumber 
\\&\quad
+24D_{n_{1}-2,n_{2}+1,n_{3}+2,n_{4}-1}^{4,4}-36D_{n_{1}-2,n_{2}+2,n_{3}+2,n_{4}-2}^{4,4}+24D_{n_{1}-2,n_{2}+3,n_{3}+2,n_{4}-3}^{4,4}\nonumber 
\\&\quad
-6D_{n_{1}-2,n_{2}+4,n_{3}+2,n_{4}-4}^{4,4}+4D_{n_{1}-1,n_{2},n_{3}+1,n_{4}}^{4,4}-16D_{n_{1}-1,n_{2}+1,n_{3}+1,n_{4}-1}^{4,4}\nonumber 
\\&\quad
+24D_{n_{1}-1,n_{2}+2,n_{3}+1,n_{4}-2}^{4,4}-16D_{n_{1}-1,n_{2}+3,n_{3}+1,n_{4}-3}^{4,4}+4D_{n_{1}-1,n_{2}+4,n_{3}+1,n_{4}-4}^{4,4}\nonumber 
\\&\quad
+4D_{n_{1},n_{2}+1,n_{3},n_{4}-1}^{4,4}-6D_{n_{1},n_{2}+2,n_{3},n_{4}-2}^{4,4}+4D_{n_{1},n_{2}+3,n_{3},n_{4}-3}^{4,4}-D_{n_{1},n_{2}+4,n_{3},n_{4}-4}^{4,4}\,.\nonumber 
\end{align}
\endgroup
\item
The coefficients with $8\leq n_{3}-n_{1}$ and $n_{2}\leq-1$ are generated recursively by
\begingroup \allowdisplaybreaks 
\begin{align}\label{label9}
D_{n_{1},n_{2},n_{3},n_{4}}^{4,4} & =D_{n_{1},n_{2},n_{3}-4,n_{4}-4}^{0,0}+4D_{n_{1},n_{2}+1,n_{3},n_{4}-1}^{4,4}-6D_{n_{1},n_{2}+2,n_{3},n_{4}-2}^{4,4}+4D_{n_{1},n_{2}+3,n_{3},n_{4}-3}^{4,4}\nonumber
\\&\quad
-D_{n_{1},n_{2}+4,n_{3},n_{4}-4}^{4,4}+4D_{n_{1}+1,n_{2},n_{3}-1,n_{4}}^{4,4}-16D_{n_{1}+1,n_{2}+1,n_{3}-1,n_{4}-1}^{4,4}\nonumber 
\\&\quad
+24D_{n_{1}+1,n_{2}+2,n_{3}-1,n_{4}-2}^{4,4}-16D_{n_{1}+1,n_{2}+3,n_{3}-1,n_{4}-3}^{4,4}+4D_{n_{1}+1,n_{2}+4,n_{3}-1,n_{4}-4}^{4,4}\nonumber 
\\&\quad
-6D_{n_{1}+2,n_{2},n_{3}-2,n_{4}}^{4,4}+24D_{n_{1}+2,n_{2}+1,n_{3}-2,n_{4}-1}^{4,4}-36D_{n_{1}+2,n_{2}+2,n_{3}-2,n_{4}-2}^{4,4}\nonumber 
\\&\quad
+24D_{n_{1}+2,n_{2}+3,n_{3}-2,n_{4}-3}^{4,4}-6D_{n_{1}+2,n_{2}+4,n_{3}-2,n_{4}-4}^{4,4}+4D_{n_{1}+3,n_{2},n_{3}-3,n_{4}}^{4,4}\nonumber 
\\&\quad
-16D_{n_{1}+3,n_{2}+1,n_{3}-3,n_{4}-1}^{4,4}+24D_{n_{1}+3,n_{2}+2,n_{3}-3,n_{4}-2}^{4,4}-16D_{n_{1}+3,n_{2}+3,n_{3}-3,n_{4}-3}^{4,4}\nonumber 
\\&\quad
+4D_{n_{1}+3,n_{2}+4,n_{3}-3,n_{4}-4}^{4,4}-D_{n_{1}+4,n_{2},n_{3}-4,n_{4}}^{4,4}+4D_{n_{1}+4,n_{2}+1,n_{3}-4,n_{4}-1}^{4,4}\nonumber 
\\&\quad
-6D_{n_{1}+4,n_{2}+2,n_{3}-4,n_{4}-2}^{4,4}+4D_{n_{1}+4,n_{2}+3,n_{3}-4,n_{4}-3}^{4,4}-D_{n_{1}+4,n_{2}+4,n_{3}-4,n_{4}-4}^{4,4}\,. 
\end{align}
\endgroup
\end{itemize}

\bibliography{refs}

\end{document}